# Is there a Vector Boson Coupling to Baryon Number?

David Bailey

*Physics Department,*

*University of Toronto, Toronto, Ontario M5S 1A7, Canada*

*E-mail: dbailey@physics.utoronto.ca*

Sacha Davidson

*Center for Particle Astrophysics,*

*301 LeConte Hall, UC Berkeley, CA 94720, USA*

*E-mail: sacha@physics.berkeley.edu*

**Abstract**

Discrepancies in measurements of the QCD running coupling constant are consistent with a new vector gauge boson coupling to baryon number with strength $\alpha_B \sim 0.1$. Upsilon decays constrain such a boson to have a mass greater than about 35 GeV. It is difficult to observe such a boson in Z decays, $e^+e^-$ annihilation, or hadronic interactions.

hep-ph/9411355  21 Nov 1994

# 1. Introduction

Baryon number (B) and the lepton family number ($L_i$) are the only exact symmetries which are not known to couple to a gauge boson. Anomaly cancellation suggests that only the combination B–L could be a gauge symmetry, but this constraint can be avoided by adding additional heavy (e.g. mirror) fermions. Many authors [1-6] have suggested extensions to the Standard Model in which B–L, B, or L are gauged symmetries, but such extensions usually predict few low energy consequences [7]. In this paper we consider bounds on a gauge boson coupled to baryon number.

Relatively good limits exist on new forces coupling to L [8] or B-L [9]. The constraints on forces coupling only to B are much weaker, except in the "fifth force" limit of very light bosons.

Since the only known fundamental particles carrying baryon number are quarks, searches for a new baryonic force must be made by studying the interactions of quarks. These are, however, dominated by their strong interactions, so new baryonic forces can only be found by looking for differences between the expected Quantum Chromodynamic colour interactions and the actual interactions of the quarks. Small discrepancies in recent measurements of the running coupling of QCD has led to wide speculation [10-13] that light gluinos might exist. In this paper we study the possibility that the discrepancy is due to a new vector force coupled to baryon number.

The simplest gauge hypothesis for baryon number is that it is a U(1) symmetry, with a "$B$" gauge boson coupling to the vector quark current with strength $g_B/3$ ($\alpha_B \equiv g_B^2/4\pi$). It would also be possible, in more complicated models, to couple a *B* boson to the axial, scalar, or pseudoscalar currents [14]. In the limit of a massless gauge boson, this vector theory would be analogous to quantum electrodynamics. There are, however, significant constraints on such a boson for masses $M_B \lesssim m_\pi$ [9, 15-21]. We therefore assume the U(1)$_B$ symmetry is broken and the *B* gauge boson is massive.

## 2. Running of $\alpha_S$

Figure 1 shows current data [22, 23] on the running of the strong coupling constant $\alpha_S(Q)$. Also shown is the expected QCD evolution for $\alpha_S(M_Z)=0.12$. There is some question as to whether the values for $\alpha_S$ extracted from low energy data are fully corrected for all non-perturbative low energy effects. At higher energies the data are consistent with QCD, but the slope of the data is less than expected. It is this slight difference which may suggest that either new coloured particles (e.g. gluinos) modify the evolution of $\alpha_S$, or that an additional force exists between quarks.

A U(1)$_B$ force would increase in strength with increasing Q in exactly the same manner as QED, and modify the observed slope of $\alpha_S(Q)$. If we naively treat the observed values for $\alpha_S$ as the sum of the true QCD coupling plus a new baryonic coupling, then we can estimate the necessary strength of the new coupling. The expected evolution for $\alpha_S(M_Z)=0.1$ and $\frac{\alpha_B(M_Z)}{3^2} = 0.02$ with $M_B=0$ is also shown in Figure 1. (The factor of "$\frac{1}{3^2}$" makes explicit that a quark has baryon number 1/3.) This value of $\alpha_B/9=0.02$ is, of course, not a precise estimate since the actual determinations of $\alpha_S$ all assume that only QCD exists. The effect of a *B* gauge boson will vary from process to process and on the experimental cuts. Since the U(1)$_B$ coupling is small, it should not appreciably change the QCD anomalous dimensions, but in any given matrix element $g_S$ and $g_B$ will not appear multiplied by exactly the same terms because the $B$ boson is colourless.

Is such a new boson consistent with other data on quark interactions? Relevant data include detailed studies of heavy vector meson decays, $Z^0$ decays, $e^+e^-$ annihilations, and high energy quark scattering.

## 3. Upsilon decays

Heavy quark vector mesons normally decay via three gluons, since two gluon decays are forbidden, and electromagnetic decays are much weaker. A gauge boson coupling to baryon number would mediate decays of heavy quark ($Q\bar{Q}$) vector mesons into light quark-antiquark pairs. The rate of vector meson decays into light quarks mediated by a virtual photon or $B$ gauge boson (see Figure 2) is, to lowest order,

$$\Gamma(V \to B^*, \gamma^* \to q\bar{q}) = N_c \frac{\Gamma(V \to \mu^+\mu^-)}{e_Q^2 \alpha^2} \sum_q \left\{ \begin{array}{c} e_Q^2 e_q^2 \alpha^2 \\ + \dfrac{2 e_Q e_q \alpha \alpha_B \, M_V^2 (M_V^2 - M_B^2)}{3^2 \left( (M_V^2 - M_B^2)^2 + \Gamma_B^2 M_B^2 \right)} \\ + \dfrac{\alpha_B^2 M_V^4}{3^4 \left( (M_V^2 - M_B^2)^2 + \Gamma_B^2 M_B^2 \right)} \end{array} \right\} \quad (1)$$

where the sum is over the light quark flavours the vector meson can decay into, $N_c$=3 is the colour factor, and $e_Q$ is the fractional charge of the heavy quark. $M_B$ and $\Gamma_B \approx N_c N_f M_B \alpha_B /9$ are the mass and decay width of the $B$ gauge boson, where $N_f$ is the number of light quark flavours the $B$ gauge boson can decay into. For clarity, the 1-$(4m_c^2/M_\Upsilon^2)$ threshold factors for charm production are not shown. The narrowest of the $b\bar{b}$ resonances, the $\Upsilon$(1S), $e_Q$=1/3, $q$=($u,d,s,c$), has a total width is $\Gamma_{tot}$=52.5±1.8 KeV, and a muonic branching ratio of B.R.($\Upsilon$(1S)→$\mu^+\mu^-$) = 2.48±0.07% [22]. The small value of $\Gamma_{tot}$ excludes any value of $\alpha_B$ for a $B$ boson satisfying the resonance condition $M_B \cong M_\Upsilon$. Off resonance ($M_B \neq M_\Upsilon$), equation (1) gives

$$\Gamma(\Upsilon(1S) \to B^*, \gamma^* \to q\bar{q}) \approx \begin{pmatrix} 4.1 \\ -2.0 \times 10^2 \dfrac{M_V^2}{M_V^2 - M_B^2} \alpha_B \\ +3.1 \times 10^4 \dfrac{M_V^4}{(M_V^2 - M_B^2)^2} \alpha_B^2 \end{pmatrix} \quad \text{(in KeV)}. \qquad (2)$$

Equation(1) implies that a very light gauge boson ($M_B << M_\Upsilon$) with coupling as small as $\alpha_B/9 = 0.005$ would completely saturate the hadronic width of the $\Upsilon(1S)$, leaving no room for any QCD mediated decays. In order to have a coupling as large as $\alpha_B/9 = 0.01$, the $B$ boson must be heavier than about 17 GeV.

Even tighter restrictions on $\alpha_B$ follow from the observed characteristics of $\Upsilon(1S)$ decays. The expected decays into 3 gluons have different characteristics from decays via a $B$ boson into a quark-antiquark pair.

First, $B$ boson mediated decays into light ($u,d,s,c$) quarks are expected be flavour independent, so approximately 1/4 of the decays should be into charm. The (slightly model dependent) observed upper limit on direct charm production from the $\Upsilon(1S)$ is only 3.4% [24]. Conservatively assuming no contribution from gluon fragmentation, this sets an upper limit to any $B$ boson mediated decays into charm. Applying equation (1) to this upper limit gives limits on the $B$ boson of $\alpha_B/9 < 0.002$ for $M_B << M_\Upsilon$, and $M_B \gtrsim 30$ GeV for $\alpha_B/9 = 0.01$.

Secondly, $B$ boson mediated decays into quark-antiquark pairs will have a 2-jet topology different from the event shapes expected for 3-gluon decays. After subtraction of the expected QED contribution, the upper limit on direct 2-jet decays of the $\Upsilon(1S)$ is 5.3% [25]. This leads to the constraints $\alpha_B/9 < 0.0015$ for $M_B \ll M_\Upsilon$, and $M_B \gtrsim 35$ GeV for $\alpha_B/9 = 0.01$. Newer data [26] could further improve these constraints.

These $\Upsilon(1S)$ data indicate that $M_B \gtrsim 35$ GeV if a $B$ gauge boson is to account for the discrepancy in $\alpha_S(Q)$. The expected evolution of $\alpha_S(Q)+\alpha'_B(Q)/9$ for $M_B=35$ GeV is also shown in Figure 1, using $\alpha_S(M_Z)=0.11$ and $\alpha'_B(M_Z)/9 = 0.01$. ($\alpha'_B$ is the effective value of the $B$ coupling including the propagator mass factor.)

## 4. Other Bounds

There are many other places where on might hope to see evidence for a $B$ boson with $\alpha_B(M_Z)/9 = 0.01$ and $M_B \gtrsim 35$ GeV. We will review some of the possibilities here, and briefly discuss why we do not get any interesting bounds.

Although there is no direct coupling between the $B$ and the Z, a $B$ gauge boson could be produced in two body $B + \gamma$ decays of the Z via quark triangle loops. Current data on exclusive (e.g. Z$\to\pi^+\pi^-\gamma$ [27]) and inclusive Z$\to\gamma$+X decays [28, 29] set limits of B.R.(Z $\to B+\gamma$) $\lesssim 10^{-4}$. There appears, however, to be some disagreement in the theoretical calculations of the amplitude for such *gauge boson $\to$ gauge boson + $\gamma$* decays (compare references [30] and [31]). We expect (following the formalism of reference [30] for Z$'\to$Z+$\gamma$ decays) the branching ratio for the decay Z $\to B+\gamma$ to be less than $10^{-5}\alpha_B/\alpha$, so the current data are consistent with $\alpha_B/9 \sim 0.01$. A more restrictive bound from the Z may come in the future from $B$ boson contributions to a "T"-violating asymmetry in 3-jet events at the SLC [32].

In addition to direct observation in Z decays, the existence of a $B$ gauge boson would modify the effective colour factors observed in Z→ 4-jets events. We would expect the observed colour factors to have contributions from both SU(3)$_{QCD}$ ($T_F/C_F$=9/4, $N_C/C_F$=3/8) and the new U(1)$_B$ ($T_F/C_F$=0, $N_C/C_F$=1). For $\alpha_B$~0.01 and $\alpha_S$~0.11, we would naively expect the observed values to be $T_F/C_F$=2.06, $N_C/C_F$=0.43. The best measurements [33] are $T_F/C_F$=2.24±0.32±0.24, $N_C/C_F$=0.58±0.17±0.23, which are easily compatible with $\alpha_B$~0.01, even without considering the suppression factor due to a finite $B$ boson mass.

Below the Z resonance, one could hope to see the $B$ boson as a peak in the total cross section for $e^+e^-\to B\to q\bar{q}$. The $B$ does not couple at tree level to electrons in the fundamental Lagrangian, but a coupling is induced at one loop via vacuum polarization diagrams that mix the $B$ with the photon [34]. We assume that at sufficiently high energies the U(1)'s are orthogonal, so the mixing amplitude is finite. One expects [14, 32] that B.R.($B\to e^+e^-$) ~ $\alpha^2/16\pi^2$. Thus even at the peak of any $B$ resonance, the contribution from $e^+e^-\to B\to q\bar{q}$ is only of the order of a few percent of the $e^+e^-\to \gamma^*\to q\bar{q}$ rate, where the uncertainty comes from the vacuum polarization integral. Such a small peak might be observable, but only with difficulty. For $\alpha_B/9$ ~ 0.01, the resonance would be quite broad ($\Gamma_B$ ~ 0.1$M_B$) and could only be observed as a peak above background by scanning an energy range larger than has been typically scanned with good relative normalization by a single experiment. (For example, a small excess of events above 56 GeV has been reported [35], but the data do not reach high enough energies to show if it is a unexpected broad peak.) Normalization uncertainties are typically 5–10% (see data of Fig. 32.11 of reference [22]), so it is difficult to search for such a peak by combining data from different experiments.

Another place to look for a new force coupling only to quarks is in hadronic interactions. The obvious channel for observation of a *B* boson would be its direct *s*-channel production and 2-jet decay $q\bar{q} \to B \to q\bar{q}$. The mass range of interest here (30 GeV $\lesssim$ M$_B$ $\lesssim$ M$_Z$) is, however, too high for ISR data [36] with negligible high mass $q\bar{q}$ scattering, and somewhat low for collider experiments with thresholds of M$_{jj}$ > 40 to 140 GeV [37-39]. The collider limits on massive *B* gauge bosons will be discussed elsewhere [14]. We only mention here that the published dijet mass limits are for M $\gtrsim$ 80 GeV [40], and the dijet data are inconsistent with any very massive *B* boson (M$_B$ > M$_Z$) with coupling large enough to make an effective contribution of $\alpha'_B$~0.01 at $Q$~M$_Z$.

Finally, a *B* gauge boson could also affect *t*–channel hadron scattering. In particular, a *B* gauge boson would mediate a colour neutral force which would produce "rapidity gap" events with no intermediate hadronization between jets. The rate of such events would be $R_{gap} \lesssim \left(\frac{\alpha_B}{9\alpha_S}\right)^2$. The observed rate [41] in deep inelastic (10< $Q^2$<100 GeV$^2$) electron-proton collisions is ~5%, which would require $\alpha_B/9$ ~0.03. The stringent limits from $\Upsilon$(1S) decays rule out any *B* bosons with a low enough mass and large enough coupling to explain these rapidity gap events. More massive bosons could contribute to rapidity events in collider experiments observed at higher $Q^2$ [42], but they could probably only be identified if they produce an observable threshold effect above $Q^2 \gtrsim M_B^2$.

This work was partially supported by the Natural Sciences and Engineering Research Council of Canada. While this work was in preparation, we learned of similar work in progress by Carone and Murayama [43].

# Figure Captions

Figure 1: The values extracted from data [22, 23] for the running coupling constant $\alpha_S(Q)$, compared with the expected evolution for QCD with $\alpha_S(M_Z)=0.12$, and QCD plus a new $U(1)_B$ gauge boson coupling to baryon number (for either $M_B=0$ or $M_B=35$ GeV, with $\alpha_S+\alpha'_B/9=0.12$ at $Q=M_Z$).

Figure 2: Heavy quark vector boson decay into light quark-antiquark pairs mediated by (a) a $B$ boson coupled to baryon number, or (b) a photon.

Figure 1

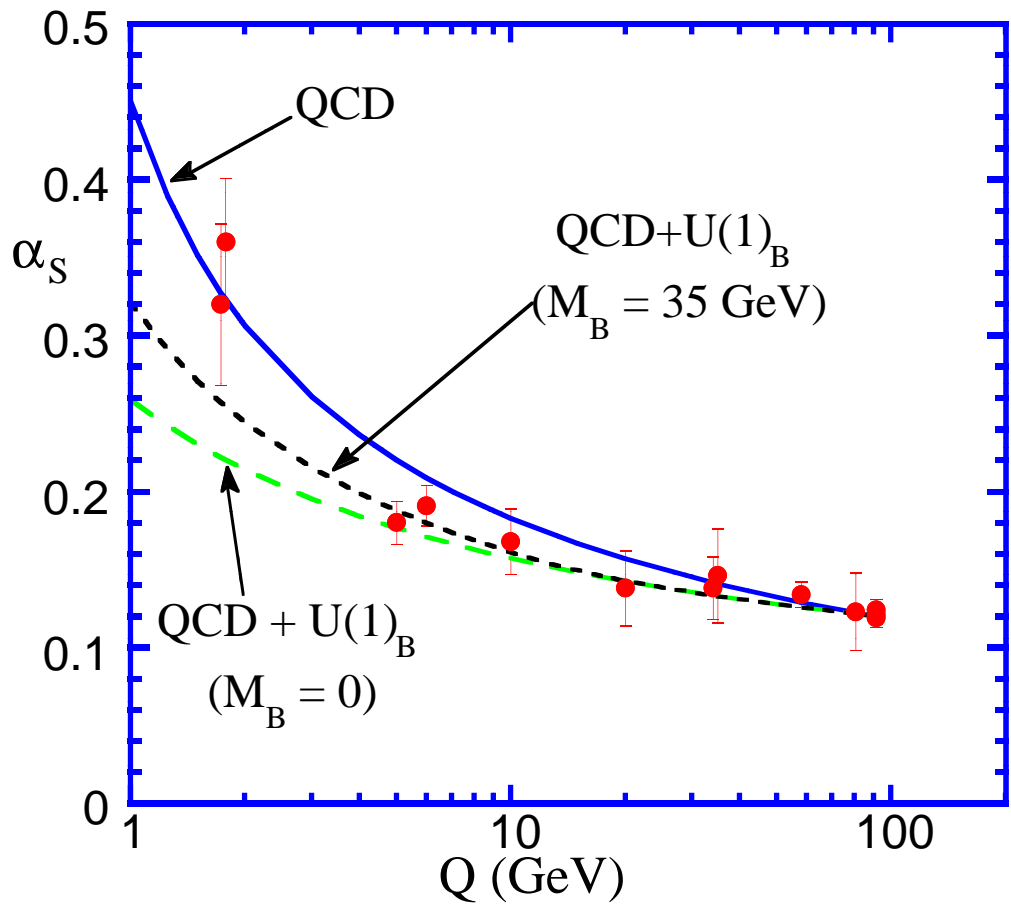

Figure 2

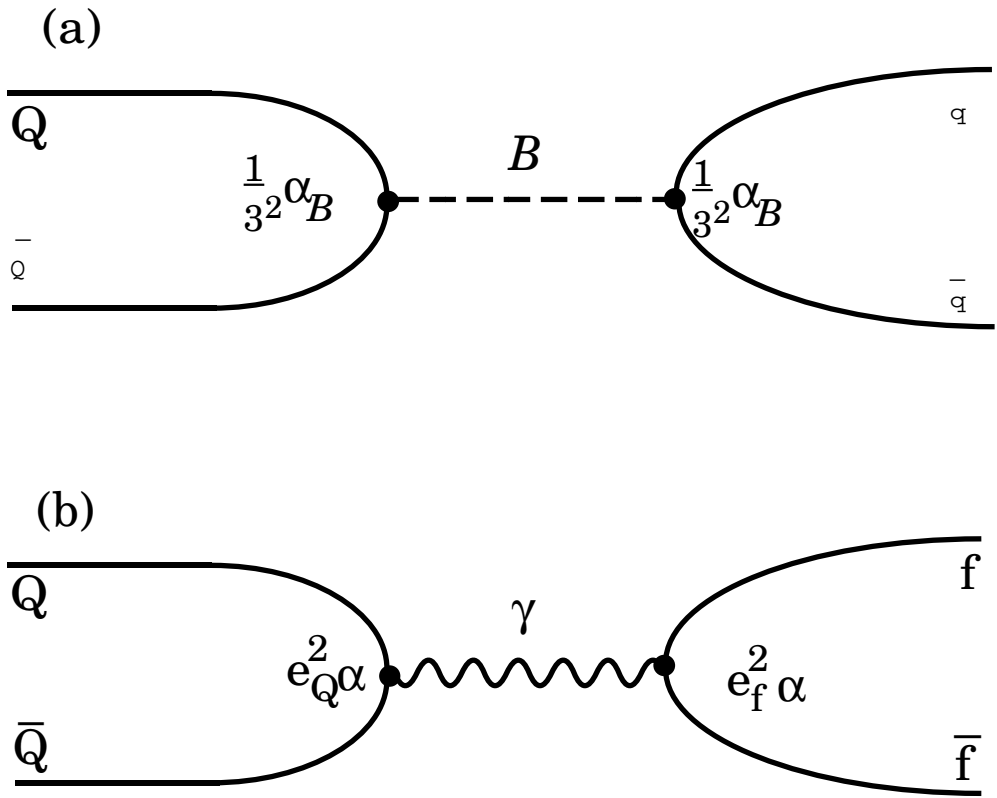